\documentclass[
 aps,
 superscriptaddress,
 prl,
 amsmath,
 amssymb,
 preprint,
]{revtex4-1}

\usepackage{graphicx}
\usepackage{dcolumn}
\usepackage{bm}
\usepackage[separate-uncertainty=true]{siunitx}
\usepackage{subcaption}
\usepackage{mhchem}
\usepackage{miller}
\usepackage{hyperref}
\usepackage{xcolor}

\usepackage[justification=justified,
            format = plain]{caption}

\makeatletter
\def\reflex#1{\ensuremath{\@millerno#1\@empty}}
\makeatother

\begin{document}

\title[Room temperature reactions of lanthanides with \ce{N2} and \ce{H2}]%
{In-situ monitoring of room temperature reactions of lanthanides with nitrogen and
hydrogen at low pressures}

\author{J.R. Chan}
\email{Jay.Chan@vuw.ac.nz}
\affiliation{The MacDiarmid Institute for Advanced Materials and Nanotechnology, School of Chemical and Physical
Sciences, Victoria University of Wellington, PO Box 600, Wellington, New Zealand.}

\author{H.J. Trodahl}
\affiliation{The MacDiarmid Institute for Advanced Materials and Nanotechnology, School of Chemical and Physical
Sciences, Victoria University of Wellington, PO Box 600, Wellington, New Zealand.}

\author{D. Lefebvre}
\affiliation{The MacDiarmid Institute for Advanced Materials and Nanotechnology, School of Chemical and Physical
Sciences, Victoria University of Wellington, PO Box 600, Wellington, New Zealand.}

\author{B.J. Ruck}
\affiliation{The MacDiarmid Institute for Advanced Materials and Nanotechnology, School of Chemical and Physical
Sciences, Victoria University of Wellington, PO Box 600, Wellington, New Zealand.}

\author{F. Natali}
\email{Franck.Natali@vuw.ac.nz}
\affiliation{The MacDiarmid Institute for Advanced Materials and Nanotechnology, School of Chemical and Physical
Sciences, Victoria University of Wellington, PO Box 600, Wellington, New Zealand.}

\date{\today}

\begin{abstract}
    The dissociative chemisorption of molecular nitrogen on clean lanthanide surfaces at ambient temperature and low pressure is explored. 
    \textit{In-situ} conductance measurements track the conversion from the lanthanide metals to the insulating lanthanide nitrides. 
    A small partial pressure of oxygen (\SI{\sim e-8}{\milli\bar}) is shown to inhibit the nitridation of lanthanides  at \SI{1e-4}{\milli\bar} of \ce{N2}. 
    The rate of nitridation as a function of nitrogen pressure is measured at low pressure for a series of lanthanide elements, gadolinium, terbium, dysprosium, ytterbium and praseodymium. 
    Exposure of the lanthanide surfaces to both \ce{N2} and \ce{H2} results in the formation of \ce{NH3}.
\end{abstract}

\maketitle

\section{Introduction}

Breaking molecular nitrogen \ce{N2} into more useful forms, i.e. nitrogen fixation, is a challenge of paramount scientific and urgent industrial importance \cite{erisman2008century, aika2012ammonia, schlogl2003catalytic}. 
Within this context the search for routes to a facile and efficient breaking of \ce{N2} and a potential energy-efficient ammonia (\ce{NH3}) synthesis at ambient temperature are among the most elusive challenges in the chemical sciences \cite{schlogl2003catalytic, howard2006many, yandulov2003catalytic}. 
The recent discovery that a single and simple surface of lanthanide (\ce{L}) metal can break molecular nitrogen at room temperature and under pressure much lower than one atmosphere is a premium advance in the field of catalytic reaction and ammonia synthesis \cite{ullstad2019breaking}. 

In this brief article we report \textit{in-situ} monitoring of room temperature reaction of ultra-pure lanthanide surfaces with nitrogen under high vacuum, typically lower than \SI{1e-7} atmospheric pressure. 
A cornerstone of our monitoring programme is that the mild conditions of the catalytic process, low temperature and high vacuum, allow continuous monitoring of the physicochemical processes taking place in real time during the course of the reaction by means of electrical conductivity measurements. 
The lanthanide nitrides (LN) form a narrow-gap insulating (semiconductor) NaCl structure with a conductivity that contrasts by many orders of magnitude with the lanthanide metals \cite{natali2013rare, ullstad2019breaking}. 
Within that scenario the in-plane conductivity of a reacting film is such that the LN fraction has negligible conductance in comparison with the remaining metal, so that to first order a reacted layer of thickness $\delta$ reduces the conductance by the ratio of $\delta$/D, for D the initial film thickness. 
The conductivity results will be displayed in terms of that equivalent reacted depth $\delta$. 
We, then, show the effect of residual oxygen on the molecular nitrogen – lanthanide reaction.
The rate of nitridation as a function of nitrogen pressure is measured for a series of lanthanide elements.
Finally, we show that experimental exposure of the lanthanide surfaces to both \ce{N2} and \ce{H2} results in the formation of \ce{NH3}.

\section{Experimental details}

Thin films of lanthanide elements were formed via electron-beam physical vapor deposition in an ultra-high vacuum chamber with a base pressure of \SI{<1e-8}{\milli\bar}. 
Sapphire (\ce{Al2O3}) \hkl(0001) substrates were held at ambient temperature, in which case the polycrystalline lanthanide films grow preferentially with the hexagonal close-packed net lying in plane; the c-axis of the metallic lanthanide was aligned along the film normal. 
The lanthanide sources (praseodymium (\ce{Pr}), gadolinium (\ce{Gd}), terbium (\ce{Tb}), dysprosium (\ce{Dy}), ytterbium (\ce{Yb})) had purities of 3N/4N. 
Electrical feedthroughs permitted \textit{in-situ} measurement of the in-plane electrical conductance during and after the deposition of a metal film and a quadrupole mass spectrometer was available for analysing the residual gas composition. 
Molecular nitrogen \ce{N2} and hydrogen \ce{H2} were introduced through separate mass flow controllers to provide working pressures between \SIrange{e-6}{e-4}{\milli\bar}. All measurements were performed at room temperature without heating of the coated surfaces inside the chamber. 

\begin{figure}[t]
    \includegraphics[width = 8.6cm]{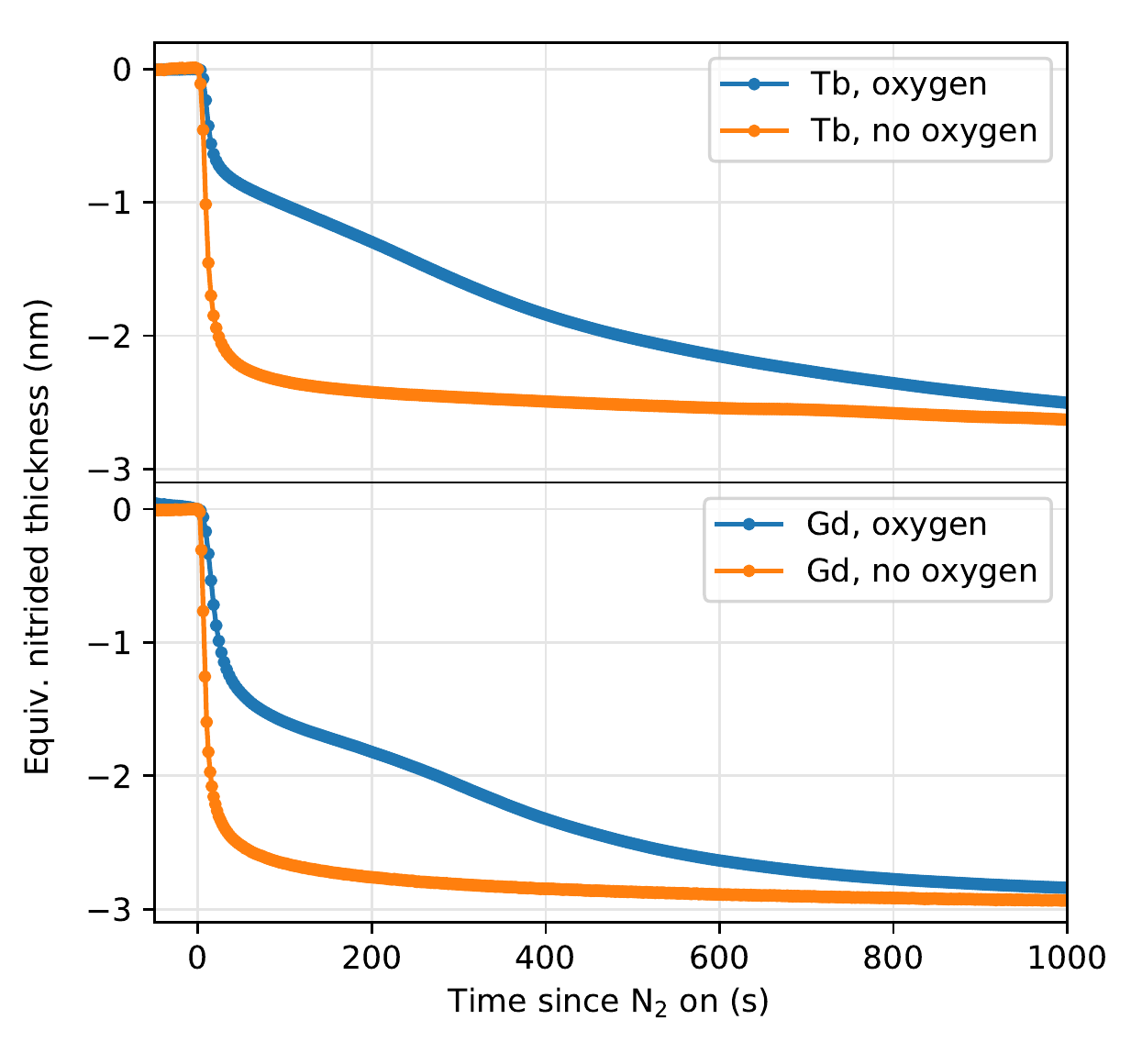}
    \caption{
        Nitridation of as-deposited \SI{20}{\nano\metre} \ce{Tb} and \ce{Gd} thin films. 
        \ce{N2} is introduced with a partial pressure of \SI{1e-4}{\milli\bar} and the corresponding conductance across the films measured. 
        An \ce{O2} partial pressure of \SI{1e-8}{\milli\bar} prevents a rapid initial decrease of the conductance compared to when no oxygen is present (\SI{<1e-10}{\milli\bar}).
    }
    \label{fig:tb_gd_nitridation}
\end{figure}

\begin{figure}[t]
    \includegraphics[width = 8.6cm]{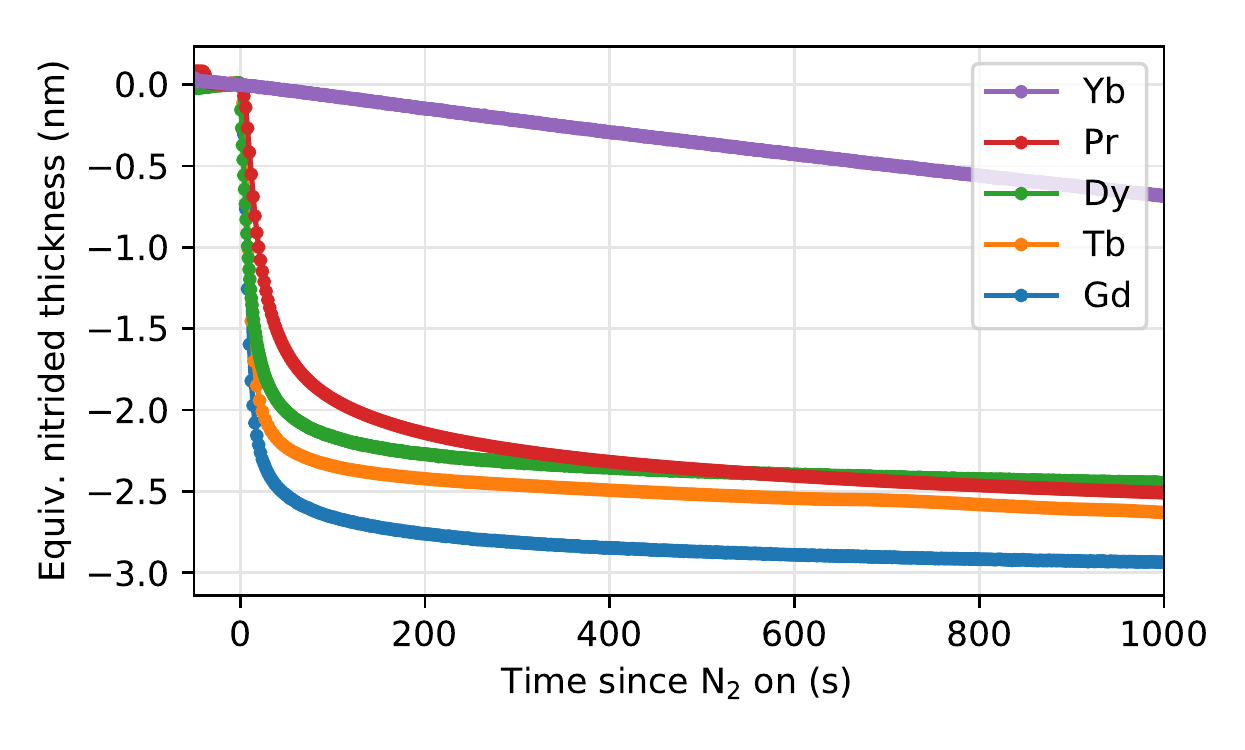}
    \caption{
        Nitridation of \ce{Yb}, \ce{Pr}, \ce{Dy}, \ce{Tb} and \ce{Gd} thin films. 
        \ce{Yb} does not react with molecular nitrogen so no sharp drop in the conductance is observed when exposed to \SI{1e-4}{\milli\bar} \ce{N2} compared to the other lanthanides.
    }
    \label{fig:lanthanide_nitridation}
\end{figure}

As a first step we report that the \ce{N2} reaction is retarded by a low concentration of \ce{O2}. 
Thus Fig.~\ref{fig:tb_gd_nitridation} compares the conductivity loss with and without a small partial pressure of \ce{O2} on \ce{Tb} and \ce{Gd} surfaces. 
\SI{20}{\nano\metre} layers of \ce{Tb} and \ce{Gd} were deposited onto a \ce{Al2O3} substrate with \textit{in-situ} electrical contacts and the resistance across the film measured during exposure to \SI{2e-4}{\milli\bar} of \ce{N2}. 
An oxygen partial pressure of \SI{\sim 1e-8}{\milli\bar}, fully four orders of magnitude smaller than the \ce{N2} pressure, is observed to limit the reaction with \ce{N2}. 
The sequence (i) rapid drop, (ii) pause, followed by (iii) a resumption of the reaction, was a feature sometimes seen also in earlier studies.\cite{ullstad2019breaking} When nitridation is performed with no residual oxygen (detection limit \SI{<1e-10}{\milli\bar}), a much stronger drop in conductance is observed in the first tens of seconds after being exposed to \ce{N2}. 
The final drop in conductance with and without \ce{O2} remains similar, corresponding to \SI{2.5}{\nano\metre} and \SI{3.0}{\nano\metre} for \ce{Tb} and \ce{Gd}, respectively, after 20 minutes.
Fig.~\ref{fig:lanthanide_nitridation} shows the nitridation of \ce{Pr}, \ce{Dy}, \ce{Tb} and \ce{Gd} thin films as measured by a drop in conductance with no \ce{O2} present. 
For comparison we show the absence of the reaction in \ce{Yb} exposed to \ce{N2}, which has been identified, along with \ce{Eu}, as not reacting with unactivated \ce{N2} \cite{richter2011electronic, binh2013europium, warring2014ybn}.

The long-term evolution of the reacted depth seen in Fig.~\ref{fig:lanthanide_nitridation} reflects the diffusion rate of N ions after they enter the lattice, which is clearly similar among the lanthanides. 
However, it is clear even at the poor resolution that there are differing initial rates, which reflect their differing reaction rates at a pure lanthanide surface. 
Those rates correspond to 0.1, 0.3, 0.25 and 0.15 nm/s for the lanthanides of increasing nuclear charge: \ce{Pr}, \ce{Gd}, \ce{Tb}, and \ce{Dy}, respectively.

\begin{figure}[t]
    \includegraphics[width = 8.6cm]{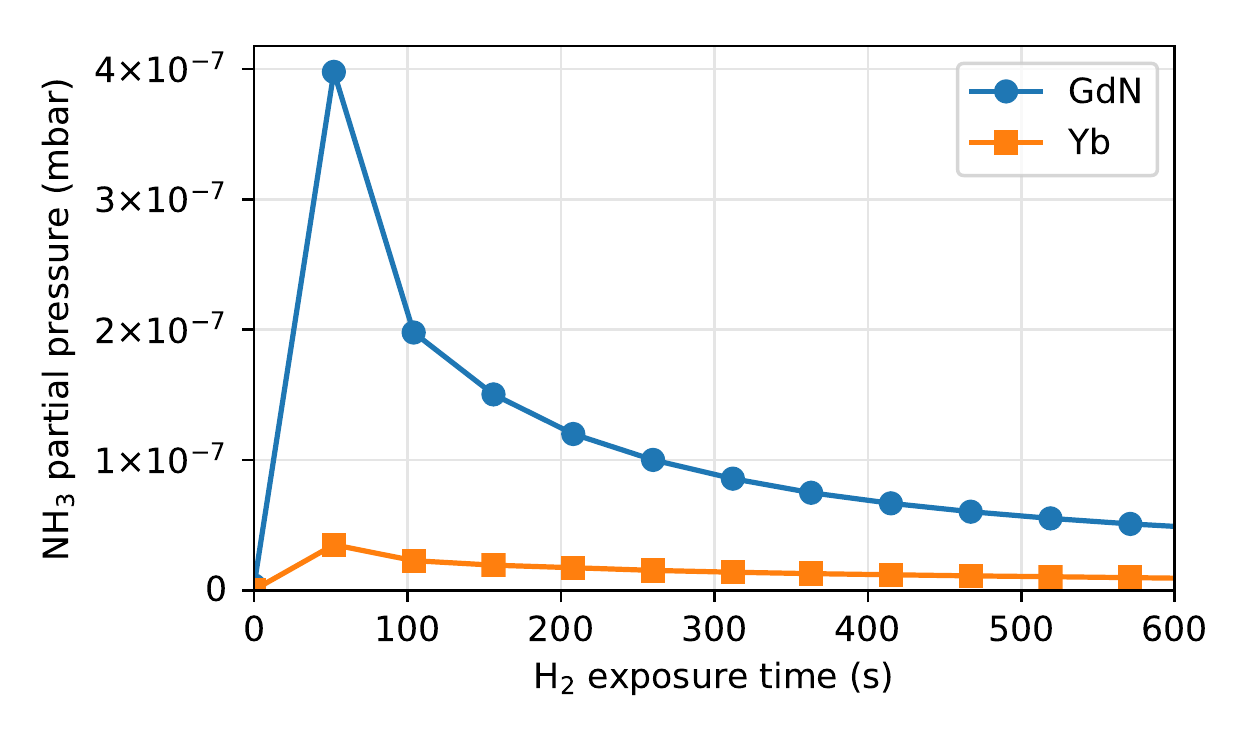}
    \caption{
        Evolution of \ce{NH3} from nitrogen-reacted \ce{GdN} (blue circles) and \ce{Yb} (orange squares) surfaces exposed to \SI{3e-4}{\milli\bar} of \ce{H2}.
    }
    \label{fig:nh3_partial_pressure}
\end{figure}

We have reported earlier a partial recovery of the conductivity when the \ce{N2} pressure was reduced, indicating a release of nitrogen near-surface layer; indeed it is that observation which suggests the potential for films' use as a catalyst. 
We thus have investigated a catalytic reaction by monitoring the presence of \ce{NH3} upon introduction of \ce{H2} following the nitridation reaction. 
Films of GdN were firstly exposed to \ce{N2} as described above, and the chamber briefly evacuated before introducing \SI{3e-4}{\milli\bar} of \ce{H2}. 
Fig.~\ref{fig:nh3_partial_pressure} shows the subsequent partial pressure of \ce{NH3} as recorded by the quadrupole mass spectrometer. 
As the \ce{H2} is introduced, an \ce{NH3} partial pressure \SI{4e-7}{\milli\bar} is detected, dropping to \SI{5e-8}{\milli\bar} after \SI{10}{\minute}, and finally to \SI{1e-8}{\milli\bar} after more than 30 minutes. 
Interestingly the time scale for the reductions correspond well to the times noted above for initial consumption of nitrogen and the subsequent diffusing in the GdN network. 
Clearly \ce{N} chemisorbed to the \ce{Gd} surface coating inside the deposition chamber is able to react with the impinging \ce{H2} and be released into the vacuum chamber. 
In contrast, \ce{Yb} exposed to \ce{N2} shows a \ce{NH3} partial pressure of \SI{2e-8}{\milli\bar}, close to the baseline of detection via the mass spectrometer. 
The catalytic activity is seen in the figure to be missing following the same procedure with \ce{Yb}, which is seen above (or in our previous paper) not to react with \ce{N2}. 

It should be noted that the baseline detection of \SI{\sim e-8}{\milli\bar} of \ce{NH3} is seen when \ce{N2} and \ce{H2} are simultaneously present in the chamber, possibly due to a reaction within the ionization chamber of the mass spectrometer. 
However this is still more than an order of magnitude lower than the ammonia level detected during the catalytic reaction with Gd. 
Furthermore, the catalytic reaction takes place over the entire coated surface inside the chamber, complicating quantitative analysis of the reaction characteristics from these measurements.

\section{Conclusions}

We have explored the dissociative chemisorption reaction of \ce{N2} and \ce{H2} with clean surfaces of lanthanide elements at ambient temperature and low pressures. 
The conversion of the lanthanide metal surface to the insulating nitride is traced using in situ conductance measurements during exposure to unactivated \ce{N2} gas.  
It is observed that a small \ce{O2} concentration can inhibit the rate of reaction of the lanthanide with \ce{N2}.
The chemisorbed \ce{N} is able to be released from \ce{Gd}/\ce{GdN} surfaces exposed to \ce{H2} in the form of \ce{NH3} as measured by mass spectrometry, also at room temperature.

\bibliography{bibliography}{}
\bibliographystyle{ieeetr}
\end{document}